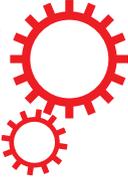

# Prediction of miRNA-disease associations with a vector space model

Claude Pasquier[1,2] & Julien Gardès[3]

MicroRNAs play critical roles in many physiological processes. Their dysregulations are also closely related to the development and progression of various human diseases, including cancer. Therefore, identifying new microRNAs that are associated with diseases contributes to a better understanding of pathogenicity mechanisms. MicroRNAs also represent a tremendous opportunity in biotechnology for early diagnosis. To date, several in silico methods have been developed to address the issue of microRNA-disease association prediction. However, these methods have various limitations. In this study, we investigate the hypothesis that information attached to miRNAs and diseases can be revealed by distributional semantics. Our basic approach is to represent distributional information on miRNAs and diseases in a high-dimensional vector space and to define associations between miRNAs and diseases in terms of their vector similarity. Cross validations performed on a dataset of known miRNA-disease associations demonstrate the excellent performance of our method. Moreover, the case study focused on breast cancer confirms the ability of our method to discover new disease-miRNA associations and to identify putative false associations reported in databases.

MicroRNAs (miRNAs) are a class of 19–24 nucleotide single-stranded non-coding RNAs that can regulate gene expression at the post-transcriptional level by binding with 3′ untranslated regions (UTRs) of the target mRNAs through base pairing. Every miRNA might regulate from a dozen to thousands of genes, and one target gene could also be regulated by hundreds of miRNAs. These miRNA-mRNA interactions play critical roles in many physiological processes, such as development, apoptosis, differentiation and metabolism. Additionally, miRNA dysregulations are closely related to the development and progression of various human diseases, including cancer.

The amount of data gathered on this topic is still relatively low at the current time. To address this limit, mathematical modeling is an appropriate solution because it allows us to focus research and development effort on high-potential new associations of miRNA with diseases.

Because miRNAs act mainly by targeting mRNAs for cleavage or translational repression[1], several modeling methods based on the characteristics of the targets were presented. From the similarity of the target mRNAs, Jiang *et al.*[2] inferred a network that represents the similarities between miRNAs. By combining this network with known mRNA-disease associations, they estimated the correlation between the miRNAs and diseases. In subsequent work, they improved the accuracy of the method by accounting for the similarities between diseases[3] and by incorporating the semantic similarity between miRNAs based on their Gene Ontology (GO) annotations[4]. Shi *et al.*[5] also exploited the knowledge of target mRNAs by working at the level of the protein-protein interaction network. They identified novel miRNA-disease associations by analyzing the links between proteins that are involved in a disease and proteins that are targeted by an miRNA. The problem with these types of approaches is that the number of experimentally verified miRNA-mRNA interactions is currently very low. To overcome this limitation, the authors used data that was generated by predictive programs that have high false-positive and false-negative rates[6].

Based on the assumption that functionally related miRNAs tend to be associated with phenotypically similar diseases[7,8], several methods have been proposed. Chen and Zhang[9] designed a method to infer miRNA-disease associations using only the phenotypic similarities between those diseases. The performance of this method is limited in that it does not account for the similarities between miRNAs.

[1]University of Nice Sophia Antipolis, I3S, UMR 7271, 06900 Sophia Antipolis, France. [2]CNRS, I3S, UMR 7271, 06900 Sophia Antipolis, France. [3]BIOMANDA, 2720 Chemin St Bernard, Les Moulins I Batiment 4, 06220, Vallauris, France. Correspondence and requests for materials should be addressed to C.P. (email: claude.pasquier@unice.fr)

Other research studies jointly use the functional similarities between miRNAs and the phenotypic similarities between diseases. Chen *et al.*[10] predicted miRNA-disease associations using a random walk with restart algorithm on a dataset composed of known miRNA-disease associations and miRNA-miRNA functional similarities. Later, they enriched their dataset with data regarding disease similarities and applied a method of regularized least squares to discover new miRNA-disease associations[11]. From an miRNA functional similarity network in which higher weights are attributed to links between miRNAs that belong to the same family or the same cluster, Xuan *et al.*[12] proposed a method based on the weighted k-most similar neighbors.Two years later, they modeled the prediction process as a random walk on the same type of network[13]. The main problem with these approaches is that the functional similarities between the miRNAs remain largely unknown. All of the algorithms presented above are based on similarity scores that are calculated with the method of Wang *et al.*[14], which estimates them from the similarities between related diseases.

The miRNA-disease network belongs to a specific class of networks (bipartite networks), whose nodes are divided into two sets, X and Y, and only connections between nodes in different sets are allowed. Recently, much attention has been given to bipartite networks, in particular, with regard to the question of how to assign weights to the links. Recommendation systems seek to predict the 'preference' or the 'contribution' from a node of one set to a node of the other set[15]. The work performed in this field can be fully applied to the prediction of the contribution of miRNAs to diseases. Li *et al.*[16] used a recommendation system to predict miRNA-disease associations. The authors reported good results but also suggested some improvements, such as integrating miRNAs and disease similarities.

Jiang *et al.*[17] presented a system that use a Support Vector Machine (SVM) to predict the associations between miRNAs and diseases. Some other studies, which are based on other supervised learning methods, can be found in the literature. However, the use of supervised methods implies having positive miRNA-disease associations but also negative miRNA-disease associations. This requirement is probably the largest drawback of such approaches.

In view of the work presented above, we can make several observations.

The first observation is that the quality of the prediction increases as the methods combine several sources of information. However, some of the biological knowledge that is available in the domain is under-used in existing programs. It is well known, for example, that members of the same miRNA family all share the same predicted targets[18]. It has also been reported that neighboring miRNAs show significant coexpression, which suggest a common regulatory pathway[7,19]. It is also obvious that knowledge regarding miRNA depicted in plain text in scientific studies contains valuable information.

The second observation is that each miRNA is associated with a set of heterogeneous data (associated diseases, targeted genes, distance from neighbors, plain text information) that makes it difficult to use traditional machine learning algorithms.

To solve these problems, we developed a method, called MiRAI, that uses distributional semantics to reveal new information that is attached to miRNAs and diseases.

Our basic approach represents distributional information on miRNAs and diseases in a high-dimensional vector space[20,21] and defines the associations between miRNAs and diseases in terms of vector similarities.

The distributional hypothesis is the basis of statistical natural language processing; it states that the meaning of words can be determined by the context in which the words are used[22]. In our case study, we can consider that miRNAs represent words, and therefore, data associated with miRNAs play the role of context. By keeping the analogy, the goal is then to use the available information on the miRNAs as a whole (context) to infer new knowledge regarding the miRNAs (words).

A vector space model is an algebraic model for representing objects as vectors. The principle is that each component of a vector is represented by a value (a weight) that should quantify the importance of a feature in the modelized object. Whole text studies use techniques such as Latent Semantic Analysis (LSA)[23] to process vectors from a vector space model. For textual data, there are many ways to calculate the weights. Among these methods, variations of the popular term frequency–inverse document frequency weighting scheme (TF-IDF), which involves multiplying the Inverse Document Frequency measure by a Term Frequency measure[21], are frequently used.

In our scenario, the vectors that represent the objects to be analyzed (miRNAs) are not homogeneous. In addition to associations between miRNAs and words extracted from plain text documents, we must also represent the relationships beween the miRNAs and diseases, targets, families or neighboring miRNAs. Links between miRNAs and diseases, targets or families consist of binary information that depends on the existence or not of an association. Neighboring relationships between miRNAs are expressed as integers that represent a distance along a chromosome in base pairs.

For textual data, for which LSA is well suited, we applied the TF-IDF weighting scheme. For other data, it was necessary to determine a weighting scheme that could represent the importance of the link between each miRNA and the associated data.

## Results

**Evaluation of the prediction performance.** To evaluate the ability of our method to predict disease-miRNA associations, a five-fold cross-validation is performed. For a specific disease *d*, the dataset is randomly partitioned into five equal sized subsets. Four of the five subsets are used to create the latent space, while the omitted subset is retained for querying and testing the model. In the testing test, all of the associations between miRNAs and *d* are removed. In addition, to be certain that the miRNAs are returned because of their underlying links with a disease and not because of some information found in the litterature, all of the data that originates from research studies are also removed. The cross-validation process is then repeated five times, with each of the five subsets used exactly once as the validation data.

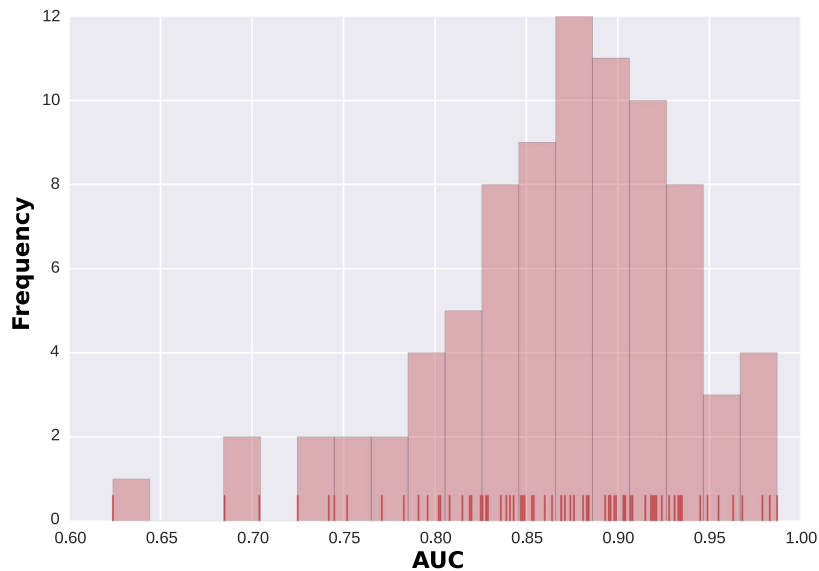

**Figure 1. Univariate distributions of AUC scores.** The small vertical ticks displayed on the horizontal axis correspond to the AUC values obtained for each of the 83 diseases. The AUC values were divided into bins of equal size. The vertical rectangles represent the number of cases for which the AUC value belong to each bin.

The latent space is then queried for the disease $d$ to obtain a ranked list of miRNAs. The higher the miRNAs associated with $d$ are ranked, the better the performance.

If an miRNA associated with $d$ has a higher rank than a given threshold $\theta$, it is considered to be a successful prediction, just as an miRNA that is not associated with $d$ that has a rank lower than $\theta$. However, because the choice of $\theta$ has a significant impact on the measures of performance, we used receiver operating characteristic (ROC)[24] analysis, which combines in a single graph the trade-off between the false positive rate (FPR) and true positive rate (TPR) over the entire range of $\theta$, where

$$TPR = \frac{TP}{(TP + FN)} \qquad FPR = \frac{FP}{(FP + TN)}$$

and *TP, TN, FP*, and *FN* are the counts of true positives, true negatives, false positives, and false negatives, respectively.

A single measure of performance is obtained by computing the area under the ROC curve (AUC) (an AUC of 1 reflects perfect classification and an AUC of 0.5 indicates random classification)[24].

MiRAI was tested on the 83 human diseases that are stored in the human miRNA-disease database (HMDD)[25], which are associated with at least 20 miRNAs.

The average AUC value obtained is 0.867 with a minimum of 0.624 for Lupus Vulgaris and a maximum of 0.987 for Hypertrophic Cardiomyopathy. Figure 1 shows the univariate distribution of the AUC scores. Figure 1 shows that few of the predictions (11 out of 82, 13%) are below 0.8. Only 2 diseases (Lupus Vulgaris and Adenoviridae Infections) obtain a score that is lower than 0.7. The scores obtained by our method are good overall. However, it is likely that the actual performance of our method is higher.

Lupus Vulgaris, which obtains the worst prediction, is associated with 62 miRNAs. In HMDD v2.0, 61 of these associations are obtained from a unique study (PMID:18998140[26]). The essential problem with this set of 61 associations is that they are collected from a study that analyzes the microRNA expression patterns in renal biopsies of lupus nephritis patients. First, the associations are deducted from raw expression values and are not confirmed. Second, the data are not related to Lupus Vulgaris. Therefore, the score obtained by our method cannot be considered to be a bad result given that the data are not accurate.

Adenoviridae Infections, the second worst prediction, is associated with 74 miRNAs, but all of them are extracted from two tables in the article PMID:20634878[27], which present the up-regulated and down-regulated miRNAs in adenovirus type 3 (AD3) infected Human laryngeal epithelial compared to controls. In this study, 14 associations are truly confirmed from Q-PCR. It is not excluded that the 60 remaining associations, based only on miRNA expression, contain false positives. By keeping the 14 verified associations only, our method obtains a more significant AUC score of 0.796. This finding might indicate that the list actually contains false associations.

For diseases that are associated with many miRNAs, the inclusion of possibly false associations has an impact on the performance. For example, the miRNAs that are associated with breast neoplasm, one of the most-studied diseases, were extracted from 319 different studies. Of the 319 studies, 16 provided more than 10 associations.

We manually checked these studies and found that for 3 of them (PMID:20099276[28] PMID:21953071[29] and PMID:22242178[30]), the extracted associations were deducted from the miRNA expression level and not confirmed. By removing 40 dubious associations that come from these studies, the AUC score for breast neoplasm jumps from 0.864 to 0.891.

| Disease name | nb assoc | MIDP | MiRAI |
|---|---|---|---|
| Carcinoma, Hepatocellular | 249 | 0.807 | 0.808 |
| Breast Neoplasms | 241 | 0.838 | 0.864 |
| Stomach Neoplasms | 207 | 0.821 | 0.815 |
| Colorectal Neoplasms | 171 | 0.845 | 0.864 |
| Melanoma | 165 | 0.837 | 0.849 |
| Lung Neoplasms | 156 | 0.876 | 0.904 |
| Heart Failure | 138 | 0.821 | 0.796 |
| Prostatic Neoplasms | 135 | 0.882 | 0.871 |
| Carcinoma, Renal Cell | 132 | 0.862 | 0.869 |
| Ovarian Neoplasms | 130 | 0.923 | 0.874 |
| Glioblastoma | 118 | 0.786 | 0.898 |
| Pancreatic Neoplasms | 118 | 0.945 | 0.928 |
| Urinary Bladder Neoplasms | 106 | 0.897 | 0.884 |
| Carcinoma, Squamous Cell | 94 | 0.870 | 0.883 |
| Leukemia, Myeloid, Acute | 77 | 0.913 | 0.895 |
| AVERAGE | | 0.862 | 0.867 |

**Table 1. Prediction results for diseases associated with the largest number of miRNAs.** The AUC scores of MiRAI are compared with the scores of MIDP, the best performing method so far.

By examining randomly some associations in HMDD v2.0, we determined that some of them concern raw expression data that is extracted from tables instead of truly confirmed associations. The inclusion of false positives in the dataset impacts all of the prediction results but more importantly the diseases that are associated with only a few miRNAs.

**Comparison with other methods.** Most of the studies that describe methods for miRNA-disease associations present the performances based on a set of well-studied diseases. To our knowledge, the state of the art is represented by MIDP[13]. In their study, Xuan *et al.*[13] compared their method to 4 other methods and showed that MIDP is the best performing method. We compared our method to MIDP for the same set of 15 human diseases that are associated with many miRNAs (Table 1). The Area Under Curve (AUC) obtained by our tool ranged from 0.796 to 0.928, with an average AUC value of 0.867. The performance is comparable to MIDP, which obtained AUC scores that range from 0.786 to 0.945, with an average of 0.862.

Because the MiRAI method is based on the similarities between miRNAs, it was expected that we would obtain good performance results for miRNAs that are associated with a large amount of data. What is more surprising is that the method perform equally well for diseases that are associated with only a few miRNAs. The average AUC value for the 27 diseases that are associated with 20 to 30 miRNAs is 0.875, with a minimum of 0.705 for Rectal Neoplasms and a maximum of 0.987 for Hypertrophic Cardiomyopathy. The prediction results for all of the tested diseases are in Supplementary Table S1.

**Identification of putative false associations.** For each disease, our method produces a list of ordered miRNAs that are potentially associated with it. The miRNAs that have a high probability to be associated with a given disease are located at the beginning of the list, while the miRNAs that are not related to the disease are placed at the end. By looking at the ordered lists obtained for each testing set, we found out that some of the known associations are located at the very end of the list. This finding is illustrated by the ROC curves that reach 100% for the false positive before all of the true positives are confirmed. Figure 2 shows the ROC curves that were obtained for Breast Neoplasms for each testing set. All of the curves show a rapid increase in the true positive rate (which corresponds to the discovery of positive samples) followed by a rather horizontal phase in which the negative samples are reported. These features are standard characteristics of ROC curves. However, the curves in Fig. 2 shows an abnormal pattern at the end, where the positive samples are discovered after most of the negative samples are reported. This finding suggests that the positive associations that are located at the very end of the predictions list might be potentially false. An miRNA-disease association is considered to be putatively erroneous if it is reported with our method with a false positive rate of above 0.85.

Systematic application of this criterion to all diseases leads to the generation of a list of 144 potential miRNAs that are falsely associated with 42 diseases (Supplementary Table S2).

**Prediction of novel miRNA-disease associations.** According to the same reasoning that is presented above, we suggest that miRNAs that are not associated with a disease that appears close to the begining of the list of candidates could be proposed as novel associations.

New miRNAs are considered to be potentially associated with a disease if the association is reported with a true positive rate of above 0.85. Systematic application of this criteria to all diseases leads to the generation of a list of 811 potential miRNAs associated with 93 diseases (Supplementary Table S3).

www.nature.com/scientificreports/

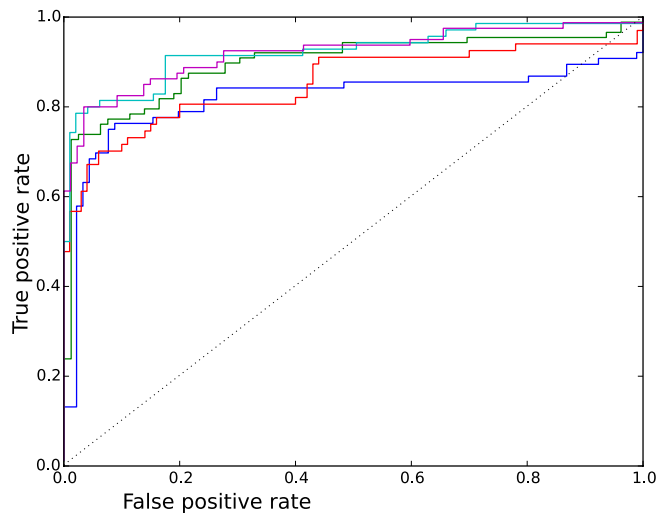

**Figure 2. ROC curves obtained at each testing set for Breast Neoplasms.** Each color line illustrates the performance of MiRAI on one partition of the data. The curve is created by plotting the true positive rate (TPR) against the false positive rate (FPR) for all threshold settings. The best possible prediction method would yield a point in the upper left corner of the ROC space, which would represent 100% sensitivity (no false negatives) and 100% specificity (no false positives). The diagonal line represents a worthless method that gives random results.

| hsa-mir-1323   | hsa-mir-1469   | hsa-mir-215  | hsa-mir-2355 |
| -------------- | -------------- | ------------ | ------------ |
| hsa-mir-3130-1 | hsa-mir-3130-2 | hsa-mir-3186 | hsa-mir-411  |
| hsa-mir-4257   | hsa-mir-4306   | hsa-mir-718  |              |

**Table 2. List of false positive miRNAs associated with Breast Neoplasms.**

**Case study: Breast Neoplasms.** To confirm the robustness of our method and its ability to discover new information on major human pathologies, we present a case study for one of the most-studied diseases in the miRNA topic: breast neoplasm.

Table 2 shows the list of miRNAs that are known to be associated with Breast Neoplasms and that appear very far in the prioritized list of miRNAs that are associated with this disease. Among them, 9 miRNAs come from a table of differentially expressed miRNAs that are presented in a publication of Schrauder et al.[30] (PMID:22242178, already discusses above). The 2 remaining miRNAs are examined in a study by van Schooneveld et al.[31] (PMID:22353773). The teams obtained results on circulating miRNAs (from blood or serum) that could be used as biomarkers to diagnose early stage breast cancer. According to Jarry et al.[32], the conclusions based on circulating miRNAs should be made with caution. The basic assumption in the studies on circulating RNA is that the most serious illness of a patient will significantly influence the quality and quantity of the miRNA found in blood. Several causes are responsive to miRNA liberation in vessels: secretions of miRNA[33], inflammatory reactions[34], cell destruction[32], and minor diseases, among others.

Inflammatory reactions are a common process in immune responses. During cancer development, chronic activations of the inflammatory response have been observed[35]. These phenomena influence the pattern of circulating miRNAs. Several articles[36–38] show a link between miR-215 (a false positive) and inflammation. These results tend to confirm that the miRNAs listed in Table 2 are more implied in the general pathways of the immune response than in the carcinogenesis phenomena of breast cancers.

Blood is a multifactorial environment, and it is impossible for a physician to diagnose all of the diseases that are present in a patient at a given time. In these conditions, only specific miRNAs for a disease can be used as biomarkers. The works of Jarry et al.[32] failed to show a specific pattern for breast cancer. The miRNAs previously quoted should be artefacts of the experiment according to our prediction method. The association of these miRNAs with Breast Neoplasms should be considered with caution.

Table 3 shows a list of 14 miRNAs that are predicted to be associated with Breast Neoplasms, whereas their associations are not reported in HMDD v2.0. Most of the associations are confirmed in the recent litterature. However, 11 associations were reported in studies that date from before the release of HMMM v2.0. This circumstance highlights the exhaustiveness limit of miRNA databases whose filling is still manual. Here, miR-208a, whose association with breast cancer is not reported in any research study, is proposed as a new miRNA that is associated with this disease.

### Discussion

A new method based on a vector space model was implemented to predict potential miRNA candidates for diseases with known related miRNAs. Our approach integrates and makes use of various sources of information that are related to miRNAs. It exhibits encouraging results and will allow us to focus research effort on interesting paths. Indeed, the costs of laboratory validations are still consequential and only a fraction of hypotheses can be humanly explored. Bioinformatics becomes the essential tool for modern research strategies.

| miRNA name | described in (PMID) | publication date |
|---|---|---|
| hsa-mir-106a | 19706389 | Sept. 2009 |
| hsa-mir-130a | 23528537 | Jun. 2013 |
| hsa-mir-138-1 | 23300839 | Dec. 2012 |
| hsa-mir-138-2 | 23300839 | Dec. 2012 |
| hsa-mir-142 | 25406066 | Nov. 2014 |
| hsa-mir-144 | 25465851 | Dec. 2014 |
| hsa-mir-150 | 24312495 | Dec. 2013 |
| hsa-mir-15b | 22908280 | Sept. 2012 |
| hsa-mir-181c | 23524334 | Jul. 2013 |
| hsa-mir-19b-2 | 21059650 | Jan. 2011 |
| hsa-mir-208a | | |
| hsa-mir-30e | 19432961 | May 2009 |
| hsa-mir-378a | 20889127 | Oct. 2010 |
| hsa-mir-99a | 21575166 | May 2011 |

**Table 3. List of new miRNAs associated with Breast Neoplasms by our method.**

Our work allowed associating new miRNAs with diseases. For diseases where the research efforts on miRNAs are less important, we suggest lists of putative candidates to target as a priority in future laboratory experiments. Some strange results were observed, which highlighted putative false associations that are stored in databases.

During cancer development, the chronic activations of the inflammatory response influence the pattern of circulating miRNAs[35]. Our method could provide a solution to this difficulty by highlighting the components that belong to this noise.

## Methods

**Data preparation.** We downloaded the experimentally verified miRNA-disease associations from the human miRNA-disease database (HMDD) v2.0 of June, 14, 2014[25]. As in previous studies[11,13,14], all of the stem-loop sequences that produce the same mature miRNA were merged into one group. Most of the miRNAs were associated with diseases that are included in the U.S. National Library of Medicine's controlled vocabulary thesaurus, the Medical Subject Headings (MeSH) terms. For the few disease names that are not listed in the MeSH, we manually mapped their names to the more relevant MeSH terms. After this pre-processing step, the dataset contained 10,737 associations between 559 miRNAs and 369 diseases. The miRNA targets were downloaded from the miRTarBase, which is a database for experimentally validated miRNA-mRNA interactions[39]. The dataset contains 48,030 associations between 1,886 miRNAs and 11,985 targets. For each miRNA, we also collected, from miRBase[40], its family and its genomic location. Abstracts of the research studies that were associated with each miRNA in miRBase were also collected.

**Creation of the matrices.** The collected data contains, for each miRNA, its known associated diseases, its target mRNAs, its family, the distance to its neighbor miRNAs and abstracts of associated studies in plain text format. This result can be represented by 5 distinct matrices, each one holding a different type of association.

**miRNA-disease associations.** The miRNA-disease associations are stored in an $m \times d$ matrix $MD = [x_{ij}]$, where $m$ is the number of miRNAs, $d$ is the number of different diseases, and $x_{ij}$ is equal to 1 if an association between miRNA at position $i$ and disease at position $j$ is present in the dataset, or is 0 otherwise. The weights of the associations are then updated by the following function: $x_{ij} = max(simil(j, k) \times x_{ik})_{k=1}^{d}$ where $simil$ is a similarity measure that accounts for the distance between two diseases in the MeSH ontology. It is calculated with the formulae proposed by Lin[41] using the implementation provided by the Semantic Measure Library[42].

$$simil(x, y) = \frac{2 \times logP(D_0)}{logP(x) + logP(y)}$$

were $P(d)$ is the probability that a randomly selected disease is equal to $d$, and $D_0$ is the most specific disease that subsumes both $x$ and $y$.

**miRNA-neighbor associations.** The miRNA-neighbor associations are stored in an $m \times m$ matrix $MN = [x_{ij}]$, where $m$ is the number of miRNAs, and $x_{ij}$ is a value that indicates the proximity of two miRNAs. Bandyopadhyay et al.[7] and Baskerville et al.[19] highlighted a significant coexpression of proximal pairs of miRNAs. They noted that beyond the 50k range, the correlation drops abruptly. To mimic this correlation in the dataset, we transform a genomic distance to a weight, expressing the correlation between two miRNAs, with a sigmoid function of the following form:

$$w(dist) = 1 - \frac{1}{1 + e^{-k(dist - dist_0)}}$$

where *dist* is the genomic distance between two miRNAs, $dist_0$ is the sigmoid midpoint and is set to $6 \times 10^5$ (a distance of 60 kb is associated with a correlation of 0.5), and k is the steepness of the curve and is set to $2 \times 10^{-4}$. This dataset contains 13682 associations between 1912 miRNAs.

**miRNA-target associations.** The raw data are stored in an $m \times t$ matrix $M = [x_{ij}]$, where *m* is the number of miRNAs, *t* is the number of different target genes, and $x_{ij}$ is equal to 1 if miRNA at position *i* targets the gene at position *j* and 0 otherwise.

The *M* matrix can be represented by a bipartite network that connects items from a set of miRNAs to items from a set of targets (Fig. 3a). Deducing the relations between the items of the same set is commonly performed using a one-mode projection. The one-mode projection produces a network that contains only nodes from the same set, where two nodes are connected when they have at least one common neighboring node from the other set. Figure 3b,c show the resulting networks after the miRNA and target projection, respectively. One-mode projections are frequently used as a similarity graph[8,43,44]. However, this transformation performed on the initial graph is quantitatively biased and produces a loss of information[45].

To overcome this problem, Zhou *et al.*[15] designed a weighting method that combines partial information given by one-mode projections to extract hidden information in the network. The method consists of two steps. First, resources associated to each node of the first set flow to nodes of the second set. When applied to Fig. 3a, the result indicates that miRNA *a* receives half of the resources of target $\alpha$ ($a = \alpha/2$), *b* receives half of the resources of $\alpha$ and a third of the resources of $\beta$ ($b = \alpha/2 + \beta/3$), and so on. Second, new resources that are associated with nodes of the second set flow back to the first set. In the example of Fig. 3a, $\alpha$ receives all of the resources that are associated with *a* and half of the resources that are associated with *b*. Node $\alpha$ is now weighted with $\alpha/2 + (\alpha/2 + \beta/3)/2$ or $3\alpha/4 + \beta/6$. The result can be represented by the following $4 \times 4$ matrix, which depicts the weighted one-mode target projection.

$$W = \begin{pmatrix} 3/4 & 1/4 & 0 & 0 \\ 1/6 & 1/2 & 1/6 & 1/6 \\ 0 & 1/4 & 1/2 & 1/4 \\ 0 & 1/4 & 1/4 & 1/2 \end{pmatrix}$$

The weights of the targets correspond to the columns in matrix *W*. Thus, the weight of node $\alpha$ is specified in the first column of *W* (3/4 of node $\alpha$ + 1/6 of node $\beta$).

It is worthwhile to notice that the matrix is not symetrical. In our example, the resources that are associated with target $\alpha$ are computed by incorporating 1/6 of the resources that are associated with target $\beta$, however, in return, the contribution of target $\alpha$ to target $\beta$ is 1/4. The MT matrix, which represents the weighted relationships between the miRNAs and targets, is calculated with the following formula.

$$MT = M.W$$

In the resulting matrix, the existing associations between an miRNA and a target are associated with a weight of $\leq 1$, while non-reported associations can be assigned a weight of $> 0$. This method allows us to generate new associations that are based on the topology of the network. Figure 3d shows the weighted bipartite graph that is obtained after the processing of the graph in Fig. 3a. Solid links represent existing relationships, with new weights specified in blue. Dotted links, with weights written in red, are new implied associations.

**miRNA-word associations.** Associations between miRNA and words used in the abstract of the associated research studies are stored in an $m \times w$ matrix $MW = [x_{ij}]$, where *m* is the number of miRNAs, *w* is the number of different words that are present in the associated documents, and $x_{ij}$ is the weight of the miRNA at position *i*. The weight is calculated using the popular TF-IDF weighting scheme[21].

**miRNA-family associations.** They are stored in an $m \times f$ matrix $MF = [x_{ij}]$, where *m* is the number of miRNAs, *f* is the number of different miRNA families, and $x_{ij}$ is equal to 1 if the miRNA at position *i* belongs to family *j*, or 0 otherwise. For this matrix only, the raw values are retained.

**Combination of matrices.** The matrices contain different types of data that are represented by real values that range from 0 to 1. A large matrix *X* that gathers all of the data available regarding miRNA is simply constructed by concatenating existing matrices.

$$X = \begin{bmatrix} MD_{(m \times d)} MN_{(m \times n)} MT_{(m \times t)} MW_{(m \times w)} MF_{(m \times f)} \end{bmatrix}$$

**Dimensionality reduction.** The goal of dimensionality reduction is to convert data of high dimensionality into data of lower dimensionality, which has the effect of bringing similar things together and pushing dissimilar things apart.

Singular Value Decomposition (SVD)[46], which is the dimensionality reduction method used in Latent Semantic Analysis, is a matrix decomposition technique that can be used to transform data in a high-dimensional space to a space of fewer dimensions. With SVD, the original matrix *X* is decomposed as a factor of three other matrices, *U*, $\Sigma$ and *V*, as follow:

$$X = U\Sigma V^T$$

where *U* is an $m \times m$ matrix, $\Sigma$ is an $m \times n$ diagonal matrix with nonnegative real numbers on the diagonal, and $V^T$ denotes the transpose of *V*, an $n \times n$ matrix.

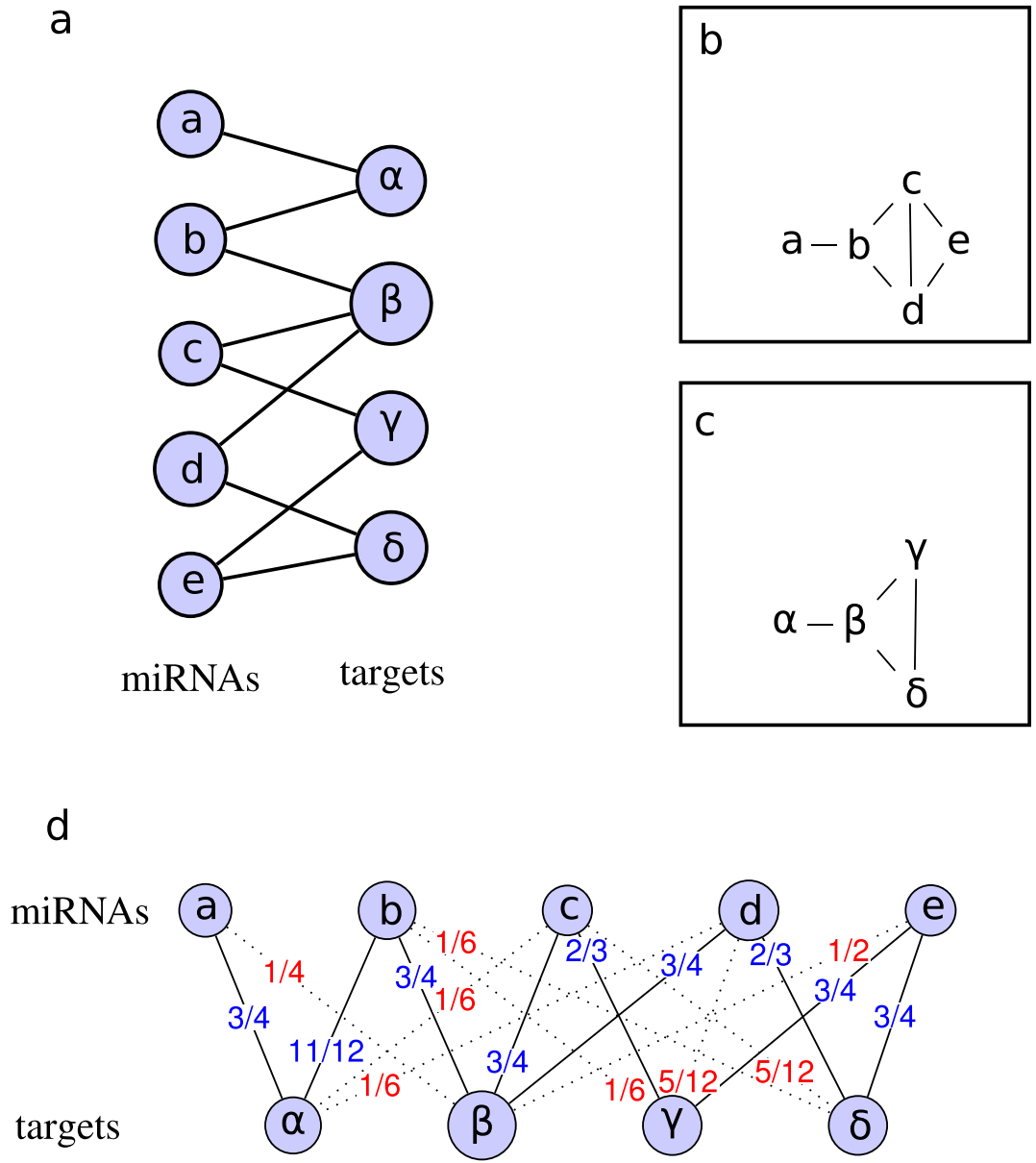

**Figure 3. Illustration of miRNA-target associations.** (**a**) miRNA-target bipartite network that connects items from a set of miRNAs to items from a set of targets. (**b**) One-mode miRNA projection that contains only nodes from the set of miRNAs. Two nodes are connected when they have at least one common connection to the same target. (**c**) One-mode target projection that contains only nodes from the set of targets. Two nodes are connected when they have at least one common connection to the same miRNA. (**d**) Weighted miRNA-target bipartite network. The graph is constructed in three steps: First, the resources that are associated with each miRNA node in graph (**a**) flow to the nodes that represent the targets. This operation indicates that miRNA *a* receives half of the resources of target $\alpha$ ($a = \alpha/2$), *b* receives half of the resources of $\alpha$ and a third of the ressources of $\beta$ ($b = \alpha/2 + \beta/3$), and so on. Second, new resources that are associated with target nodes flow back to the nodes that represent the miRNAs. In the example, $\alpha$ receives all of the resources that are associated with *a* and half of the resources that are associated with *b*. Node $\alpha$ is now weighted with $\alpha/2 + (\alpha/2 + \beta/3)/2$ or $3\alpha/4 + \beta/6$. Third, the weights are applied to the original miRNA-target network. In the new weighted graph, some of the connections that are present in the original graph appear with updated weights (black lines with weights specified in blue), while new connections are highlighted (dotted lines with weights specified in red).

It is often useful to approximate $X$ using only $r$ singular values (with $r < min(m, n)$), in such a way that we have $X = U_r \Sigma_r V_r^T + E$, where $E$ is an error or residual matrix, $U_r$ is an $m \times r$ matrix, $\Sigma_r$ is an $r \times r$ diagonal matrix, and $V_r$ is an $n \times r$ matrix. The matrix $X_r = U_r \Sigma_r V_r^T$ is the matrix of rank $r$ that best approximates the original matrix $X$. The proper choice of dimensionality is important. However, there is no general method for finding the optimal dimension. In this work, we determined the size by simulation. We tested the performance of the predictions for all of the dimensions between 50 and 500 with a step of 50 and found that using 400 dimensions was the choice that allows us to obtain better predictions. According to this result, Singular value decomposition (SVD) is applied on the combined matrix, and only the 400 first vectors are retained ($r = 400$).

**Querying.** In $X_r$, diseases and miRNAs are represented by vectors of length 400. Following the classical approach that is used in document indexing, the relatedness of two vectors is measured by their cosine distance. To obtain a ranked list of the miRNAs that are related to a disease $d$, queries are performed in the latent space by prioritizing the miRNA vectors by their distance to the vector of $d$.

## Author Contributions

C.P. and J.G. conceived the study. C.P. developed the method, C.P. and J.G. conducted the computational analysis, and C.P. and J.G. wrote the manuscript.

## Additional Information

**Supplementary information** accompanies this paper at http://www.nature.com/srep

**Competing financial interests:** The authors declare no competing financial interests.

**How to cite this article**: Pasquier, C. and Gardès, J. Prediction of miRNA-disease associations with a vector space model. *Sci. Rep.* **6**, 27036; doi: 10.1038/srep27036 (2016).